# On the Correlation between Energy Spectra and Element Abundances in Solar Energetic Particles

**Donald V. Reames**

Institute for Physical Science and Technology, University of Maryland, College Park, MD 20742-2431 USA, email: dvreames@gmail.com

**Abstract** In solar energetic particle (SEP) events, the physical processes of both shock acceleration and scattering during transport can cause energy-spectral indices to be correlated with enhancement or suppression of element abundances versus mass-to-charge ratios $A/Q$. We observe correlations for those "gradual" SEP events where shock waves accelerate ions from the ambient coronal plasma, but there are no such correlations for "impulsive" SEP events produced by magnetic reconnection in solar jets, where abundance enhancement in different events vary from $(A/Q)^{+2}$ to $(A/Q)^{+8}$, nor are there correlations when shock waves reaccelerate these residual impulsive ions. In these latter events the abundances are determined separately, prior to the accelerated spectra. Events with correlated spectra and abundances show a wide variety of interesting behavior that has not been described previously. Small and moderate gradual SEP events, with relative abundances typically depending approximately upon $(A/Q)^{-1}$ and the spectra upon energy $E^{-2.5}$, vary little with time. Large SEP events show huge temporal variations skirting the correlation line; in one case O spectra vary with time from $E^{-1}$ to $E^{-5}$ while abundances vary from $(A/Q)^{+1}$ to $(A/Q)^{-2}$ during the event. In very large events, streaming-limited transport through proton-generated resonant Alfvén waves flattens the spectra and enhances heavy ion abundances prior to local shock passage, then steepens the spectra and reduces enhancements afterward, recapturing the typical correlation. Systematic correlation of spectra and element abundances provide a new perspective on the "injection problem" of ion selection by shocks and on the physics of SEP acceleration and transport.

Keywords: Solar energetic particles · Solar system abundances · Coronal mass ejections · Shock acceleration

## 1. Introduction

The processes of acceleration and transport of solar energetic particles (SEPs) can depend upon particle velocity and upon their magnetic rigidity, which controls deflection and scattering by magnetic fields or Alfvén waves, for example. What is often called particle "energy," $E$, quoted in MeV amu$^{-1}$, actually measures velocity $E = \mathcal{E}/A = M_u(\gamma - 1) \approx \frac{1}{2} M_u \beta^2$, where $\mathcal{E}$ is the total kinetic energy, $A$ is the atomic mass, $M_u = m_u c^2 = 931.494$ MeV, $\gamma = (1- \beta^2)^{-1/2}$, and $\beta = v/c$ is the particle velocity relative to the speed of light [$c$]. The magnetic rigidity, or momentum per unit charge, is $P = pc/Q\mathrm{e} = M_u \beta\gamma\, A/Q$ in units of MV. When ions are accelerated from solar coronal plasma by a shock wave, for example, scattering back and forth across the shock depends upon magnetic rigidity, but ions gain an increment of velocity on each transit. The ions approach equilibrium power-law energy spectra, while different ion species of the same velocity reach an equilibrium pattern of enhancement or suppression, relative to their source abundances, that depends upon $A/Q$. Frequently the equilibrium energy spectrum and the pattern of $A/Q$ are both approximately power laws. Positive and negative power-law abundance patterns in large SEP events were first reported by Breneman and Stone (1985) for the elements with atomic numbers $6 \leq Z \leq 30$, and these powers of $A/Q$ have been further studied extensively for impulsive (e.g. Reames, Cliver, and Kahler, 2014a, 2014b; Reames, 2019a) and gradual (Reames, 2016, 2019b) SEP events and compared for both (Reames, 2020a). However, the relationship between powers of $A/Q$ and energy spectral indices has been briefly introduced only recently (Reames, 2020b).

If shock waves were planar structures as often assumed, ions with shorter scattering mean free paths $\lambda$ would be accelerated faster, traversing the shock more often. However, shocks can be quite complex. The review article by Jones and Ellison (1991) discusses the reduced rigidity dependence seen at the Earth's bow shock where shock smoothing allows ions with longer mean free paths to compensate by seeing larger velocity differences, as shown by Monte Carlo calculations (Ellison, Möbius, and Paschmann, 1990). Shocks are now known to be complex surfaces, modulated by waves (e.g. Trotta et al., 2020), and small-scale variations in interplanetary shocks have been observed by the *Cluster* spacecraft (e.g. Kajdič et al., 2019). Furthermore, acceleration of SEPs in magnetic reconnection sites need not involve shocks at all, and we find their spectra and abundances to be unrelated.

For many years we have recognized that two processes and locations of SEP acceleration produce events that are historically called "impulsive" and "gradual" (e.g. Reames, 1988, 1995b, 1999, 2013, 2015, 2017, 2020a, 2021; Gosling, 1993). In the small "impulsive," SEP events (Mason, 2007; Bučík, 2020), acceleration is now understood to also occur at open magnetic-reconnection sites in solar jets (Kahler, Reames, and Sheeley, 2001; Bučík et al., 2018a, 2018b; Bučík, 2020) where SEPs easily escape. Particles similarly accelerated in flares are trapped on closed magnetic loops; being unable to contribute to SEPs in space, they plunge into the loop footpoints to produce X-rays and hot, bright plasma. Impulsive events were first identified long ago by huge enhancements of $^3$He such as $^3$He/$^4$He = 1.52 ± 0.10 (e.g. Serlemitsos and Balasubrahmanyan, 1975) compared with $5 \times 10^{-4}$ in the solar wind, probably enhanced by resonant wave-particle reactions (e.g. Temerin and Roth, 1992). Element abundances in impulsive SEP events, relative to their coronal values, have now been observed to increase by a factor of ≈1000 across the periodic table (Reames, 2000; Mason et al., 2004; Reames and Ng, 2004; Reames, Cliver, and Kahler, 2014a, 2014b), varying on average as $(A/Q)^{3.64 \pm 0.15}$ (Reames, Cliver, and Kahler, 2014a) from H and He up to Au or Pb, with $A/Q$ determined at a temperature $T \approx 3$ MK in impulsive SEP events (Reames, Meyer, and von Rosenvinge, 1994; Reames, Cliver, and Kahler, 2014a, 2014b; Reames, 2019a). These power-law element abundance enhancements appear to be a direct consequence of magnetic reconnection (Drake et al., 2009).

In the larger "gradual" SEP events (Lee, Mewaldt, and Giacalone, 2012; Desai and Giacalone, 2016), particles are accelerated to much higher energies at shock waves, driven by fast, wide coronal mass ejections (CMEs) as shown in the observations (Kahler et al., 1984; Mason, Gloeckler, and Hovestadt, 1984; Reames, Barbier, and Ng, 1996; Cliver, Kahler, and Reames, 2004; Gopalswamy et al., 2012; Cohen et al., 2014; Kahler, 2001; Kouloumvakos et al., 2019) and are described by theories and models (Lee, 1983, 2005; Zank, Rice, and Wu, 2000; Ng and Reames, 2008; Afansiev, Battarbee, and Vainio, 2016; Hu et al., 2017, 2018). Many gradual SEP events show high-energy spectral breaks (e.g. Mewaldt et al., 2012); we consider only energies below these spectral breaks. Most gradual events involve acceleration of ambient coronal material (Reames, 2016, 2019b, 2020a). Comparisons of average SEP ion abundances with corresponding photospheric abundances show a characteristic dependence on the first ionization potential (FIP) of the ions that is different from that of the solar wind (Mewaldt et al., 2002; Desai et al., 2003; Reames, 2018, 2020a; Laming et al. 2019); SEPs are *not* accelerated solar wind.

Ion transport after acceleration may also involve scattering that depends upon a power of the ion magnetic rigidity (Parker, 1963), creating abundances that have a power-law dependence upon *A/Q* for ions compared at a constant velocity. For the largest gradual SEP events, scattering may even be induced by self-generated Alfvén waves (Stix, 1962, 1992; Melrose 1980) produced or amplified by the streaming SEP protons themselves (Lee, 1983, 2005; Ng and Reames, 1994, 1995; Ng, Reames, and Tylka, 1999, 2001, 2003, 2012). Thus, for example, since Fe scatters less than O as ions stream outward, Fe/O is enhanced early in an event and is therefore depleted later, creating a power-law dependence upon *A/Q* that increases early in the event and decreases later. This can become a dependence upon longitude because of solar rotation.

The clear distinction between the acceleration mechanisms in impulsive and gradual SEP events, magnetic reconnection and shock acceleration, becomes blurred when local shock waves at CMEs in jets become fast enough to reaccelerate the impulsive ions pre-accelerated in the magnetic reconnection. Also, large pools of impulsive suprathermal ions can collect from many jets, producing a $^3$He-rich, Fe-rich background that is commonly observed (Desai et al., 2003; Bučík et al., 2014, 2015; Chen et al., 2015). These pools of impulsive suprathermal seed ions are sometimes encountered by shock waves and may be preferentially accelerated (Desai et al., 2003; Tylka et al., 2005; Tylka and Lee, 2006) so they even dominate SEP abundances from weaker shocks or quasi-perpendicular shocks that favor the boosted injection velocity. All fast shock waves probably accelerate some impulsive suprathermal ions, if they happen to be available, but for many strong shocks these are only a very small fraction of the resulting SEPs (Mason, Mazur, and Dwyer, 1999).

Reames (2020a) has identified four paths to produce SEP element abundances:

(i) SEP1: "pure" impulsive events from jets, sans shock acceleration,

(ii) SEP2: impulsive events with reacceleration by a local shock,

(iii) SEP3: gradual events dominated by pre-accelerated impulsive ions at $Z > 2$,

(iv) SEP4: gradual events dominated by seed ions from ambient coronal plasma.

The SEP measurements we use in this article come from the *Wind* spacecraft near Earth. These include the relative abundances of the elements H, He, C, N, O, Ne, Mg, Si, S, Ar, Ca, and Fe, and also groups of heavier elements up to Pb from the *Low-Energy Matrix Telescope* (LEMT; von Rosenvinge et al., 1995) on *Wind*. Abundances are primarily from the 3.2–5 MeV amu$^{-1}$ interval on LEMT, although H is only available near 2.5 MeV amu$^{-1}$. Abundance enhancements are measured relative to the average SEP abundances listed by Reames (2021; see also Reames, 1995a, 2014, 2020a). We also reference event numbers from the list of gradual SEP events published in Reames (2016), and the list of impulsive SEP events published in Reames, Cliver, and Kahler (2014a).

For the SEP events we analyze, obtaining spectral indices by least-squares fitting is quite straightforward and fitted spectra are shown with most events. Obtaining power-law fits of abundance enhancements versus $A/Q$ is slightly more involved but has now been described in many articles (e.g. Reames, Cliver, and Kahler, 2014b, 2015; Reames, 2016, 2019a, 2019b, 2019c, 2020a), and has been described in detail in a review (Reames, 2018b) and a textbook (Reames, 2017, 2021). Gradual events are divided into 8-hour time intervals to obtain adequate statistics on at least the 12 main elements and find enhancements of each element, relative to previously-determined reference coronal abundances (e.g. Reames, 1995a, 2014, 2017, 2018a, 2020a). An array of 9 possible source-plasma temperatures, spaced from 0.8 to 6.0 MK, is chosen and each temperature determines values of $Q$ for each element (e.g. Post et al., 1977; Mazotta et al., 1998) allowing enhancements versus $A/Q$ to be fitted at each $T$. A minimum value for $\chi^2/m$ of fits versus $T$ determines the best-fit temperature and enhancement versus $A/Q$ which are shown for each time interval in the figures. Since the behavior of H and He is often complex, fits are obtained using the elements $6 \leq Z \leq 56$ and these are then extrapolated to protons at $A/Q = 1$. For most gradual SEP events the proton abundances are well-predicted by this extrapolated fit line (the SEP4 events), but gradual SEP events with impulsive seed particles (SEP3 events) have $T > 2$ MK and there is a large (ten-fold) proton excess above the fit line which is interpreted as an additional contribution of protons accel-erated from the ambient solar plasma in addition to the impulsive seeds (Reames, 2019b, 2019c, 2020a). This ability to determine source plasma temperatures and best-fit abundance enhancements versus $A/Q$ has been an enormous aid in the progress of SEP studies, since direct measurements of $Q$ (e.g. Klecker, 2013) are difficult and are rarely available.

The purpose of this article is to explore relationships between energy spectral indices of typical ions such as He, O, and Fe, and enhancement or suppression of element abundances versus $A/Q$, at constant velocity, relative to coronal abundances, under the approximation of power-law spectral and $A/Q$ dependence. For which SEP events do relationships exist? Why or why not? If spectra and abundances are related, when are they controlled by acceleration and when by interplanetary transport? In this poorly explored regime, evidence of no coupling between spectra and abundances can also be strong evidence about the nature of the particle source. In Sections 2 – 5, we examine behavior in several types of individual SEP events of varying size and complexity and then in Sections 6 and 7 we consider distributions of gradual and impulsive events, respectively.

## 2. Moderate-Sized Gradual SEP4 Events

Recently, Reames (2020b) suggested that properties of small- and modest-sized gradual SEP events were generally controlled by the particle acceleration at shock waves with little modification of the events by scattering during transport of the ions out to 1 AU. In that preliminary study it was found that when abundance enhancements vary as $(A/Q)^x$ and spectra vary as $E^y$, we might expect $y = x/2 - 2$.
Figure 1 shows the analysis of an event that obeys these expectations, the event of 24 August 1998. The figure shows details of the power-law fits (Figs. 1c, 1d, and 1e) and comparison of spectral indices of O with Fe and with powers of $A/Q$ for the event (Figs. 1f and 1g). This event has similar spectral indices for O and Fe (Figs. 1d, 1e, and 1f), except for some slight degradation of the power law of the highest-energy Fe point at the latest time (Fig. 1e). The process of minimization of $\chi^2/m$ versus $T$ to determine temperatures and produce Fig. 1c for this event was shown in Fig. 2 of Reames (2019b). For the event in Fig. 1, the spectral indices of Fe and O agree (Fig. 1f) and they change little with time, suggesting that they were determined early; subsequent activity by the weakening shock, seen by the peak at shock passage (Fig. 1a), only reaccelerated the same ions, maintaining their initial spectral shape.

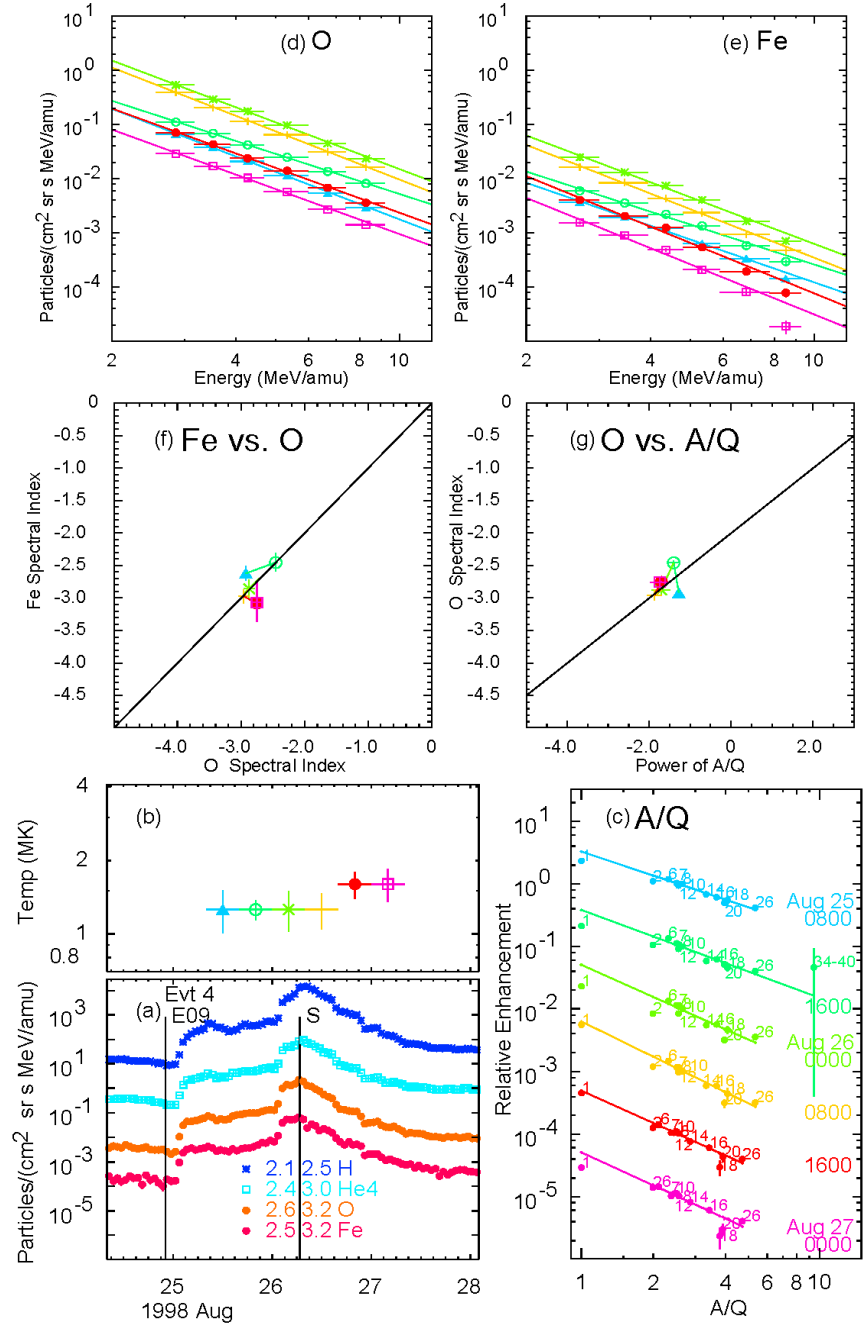

**Figure 1** (**a**) Intensities, for listed ions and energies in MeV amu$^{-1}$ and (**b**) derived source temperatures are shown versus time for gradual SEP Event 4 (list in Reames, 2016), on 24 August, 1998. Fits are shown for (**c**) enhancements of elements, listed by $Z$, versus $A/Q$, and for energy spectra of (**d**) O and (**e**) Fe. Correlation plots are shown for spectral indices of (**f**) Fe versus O and of (**g**) O versus $A/Q$. Colors for time intervals correspond in (**b**), (**c**), (**d**), (**e**), (**f**), and (**g**). In (**f**) the solid line is diagonal, y = x, in (**g**) it is y = x/2 − 2.

Similar behavior is seen in Fig. 2 for the 13 September, 2005 gradual SEP event. For this event the spectra of Fe are increasingly somewhat steeper than those of O (Fig. 2f), although the power of $A/Q$ is reasonably consistent with the O spectral index (Fig. 2g). The Fe and O measurements here span similar velocity intervals so that Fe typically has over twice the rigidity (Fig. 2c) of O. The power-law abundances are as sensitive to Fe as they are to O.

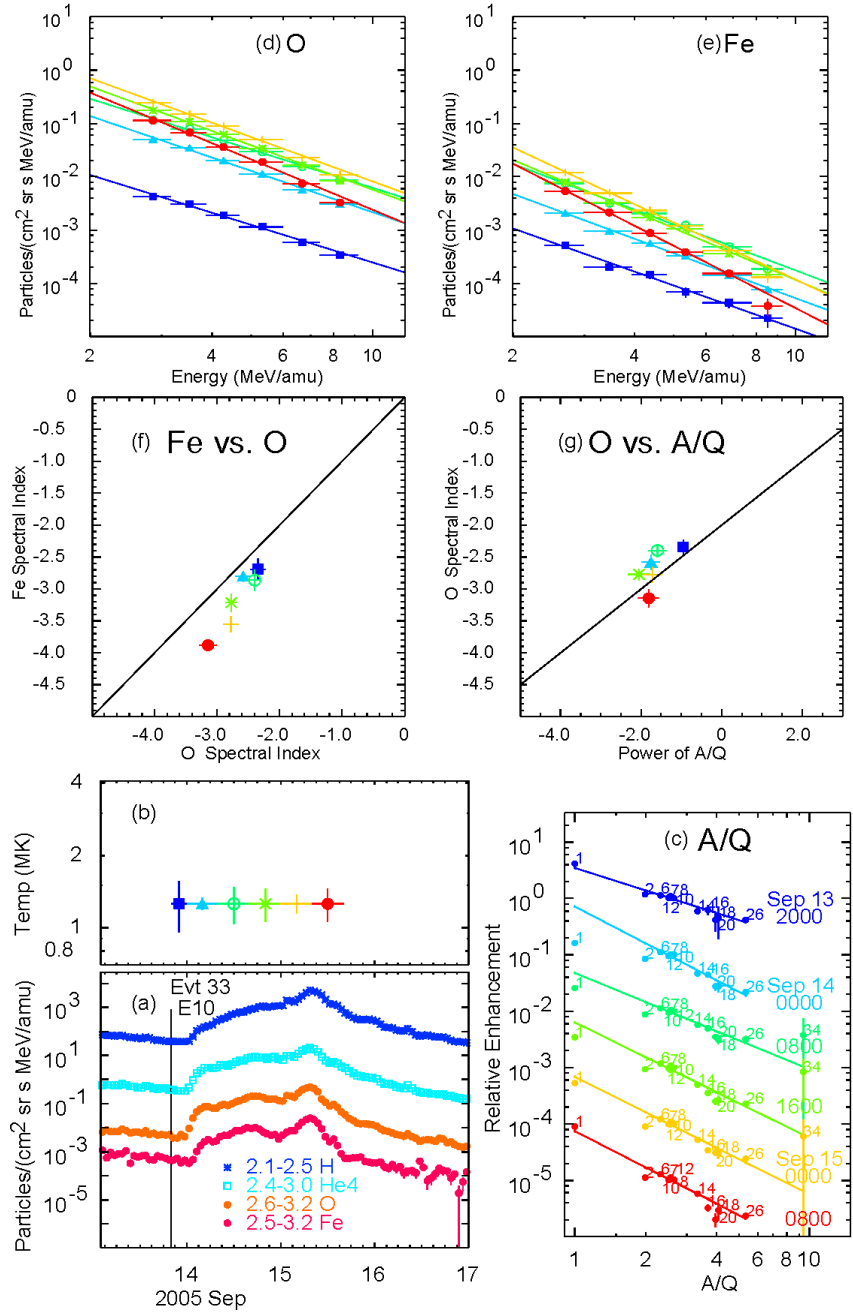

**Figure 2** (a) Intensities and (b) derived source temperatures are shown versus time for gradual SEP Event 33, of 13 September, 2005. Fits are shown for (c) enhancements of elements, listed by $Z$, versus $A/Q$, and for energy spectra of (d) O and (e) Fe. Correlation plots are shown for spectral indices of (f) Fe versus O and of (g) O versus $A/Q$. Colors for time intervals correspond in (b), (c), (d), (e), (f), and (g). In (f) the solid line is diagonal, y = x, in (g) it is y = x/2 – 2.

A third example in this event category is shown in Fig. 3. Here measurements in the first and last time periods are somewhat more erratic.

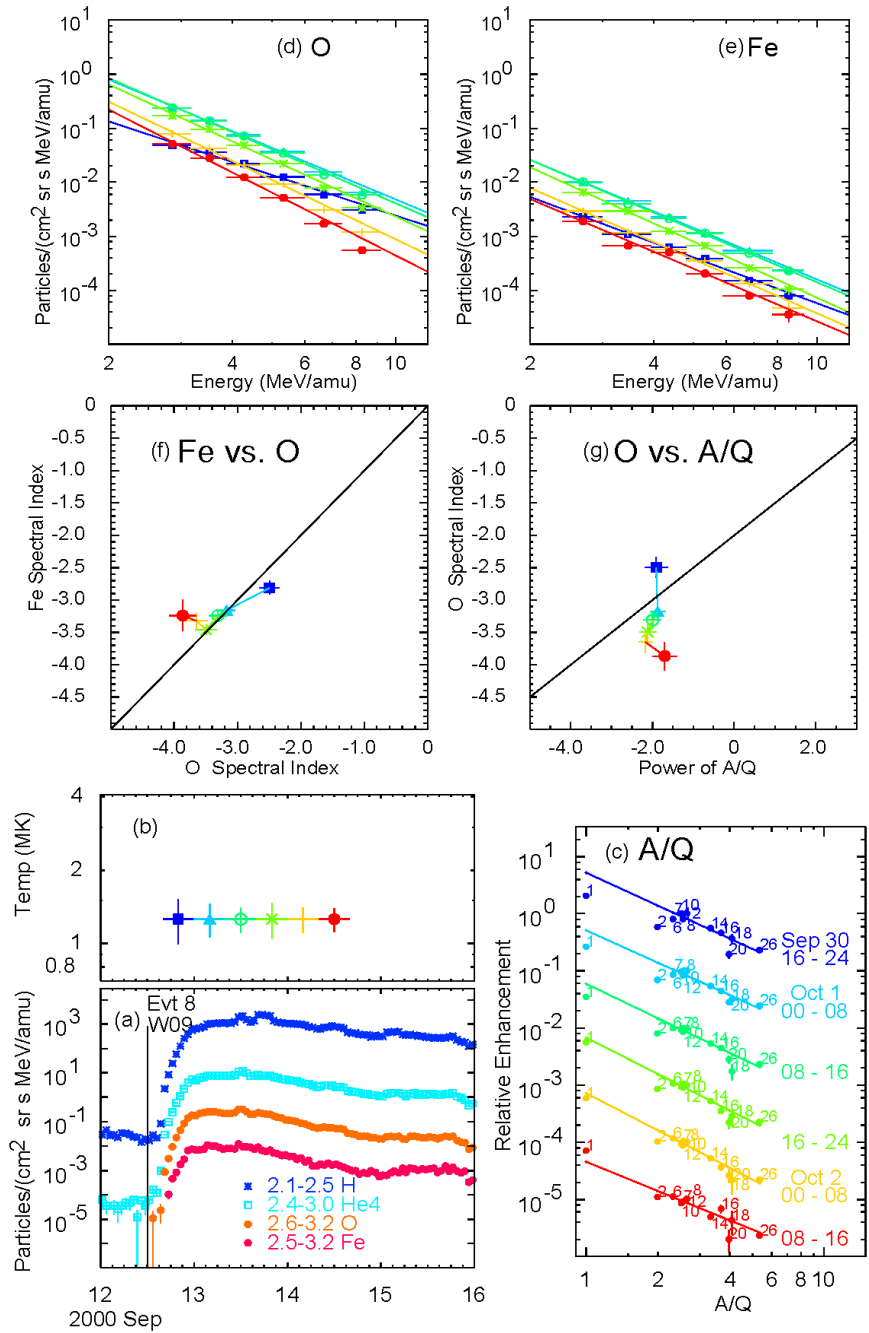

**Figure 3** (**a**) Intensities and (**b**) derived source temperatures are shown versus time for gradual SEP Event 8, of 12 September, 2000. Fits are shown for (**c**) enhancements of elements, listed by $Z$, versus $A/Q$, and for energy spectra of (**d**) O and (**e**) Fe. Correlation plots are shown for spectral indices of (**f**) Fe versus O and of (**g**) O versus $A/Q$. Colors for time intervals correspond in (**b**), (**c**), (**d**), (**e**), (**f**), and (**g**). In (f) the solid line is diagonal, y = x, in (**g**) it is y = x/2 − 2.

Thus, gradual SEP events in this category show reasonable agreement in the spectra of He (not shown), O, and Fe, and abundance patterns and spectra that do not change much during the course of an event.

# 3. SEP4 Events with Strong Evolution of Spectra and Abundances

Events in this category are large SEP4 events that show extreme changes in spectra and abundances with time during the events. Typical events are shown by Event 40 (on the list of Reames, 2016), 29 September, 2013, in Fig. 4 and the much more intense Event 23, 2 November, 2003 in Fig. 5.

These events show rapid and extensive steepening of the energy spectra, often more in O than in Fe, probably because Fe has a higher rigidity. Figs. 4d and 4e show O spectra steepening from $E^{-1}$ to $E^{-5}$ during the event. Misalignment and suppression of protons in Fig. 5c suggests that there was some instrument saturation of the proton intensity near the time of shock passage but other ions were not affected.

These events also show the classic evolution in the power of *A/Q* from +1 to -2 during the events as seen in Fig. 4c and 5c. This is usually explained in terms of transport: if Fe scatters less than O then Fe/O will be enhanced early in an event and depleted later; spectra steepen when low rigidity ions are trapped by waves while higher-rigidity ions leak away. Early and late in an event we see these leaking low-rigidity ions; nearer the shock we see the trapped low-rigidity ions. Alternatively, it would also be possible to speculate that for these magnetically well-connected events at W33 and W56, the observer is well-connected to the nose of the shock early, but that magnetic connection then sweeps around to weaker eastern flank of the shock with time where the accelerated spectra are steeper. Is this a temporal variation, a spatial variation swept past, or both? Unfortunately, diffusion theory is so flexible that we cannot tell, since it includes arbitrarily adjustable power-laws in rigidity for diffusion coefficients that may or may not actually exist. Such issues could be resolved by multiple appropriately-spaced spacecraft, but STEREO spacecraft, for example, were too widely separated to sample each longitude a few days apart during an event.

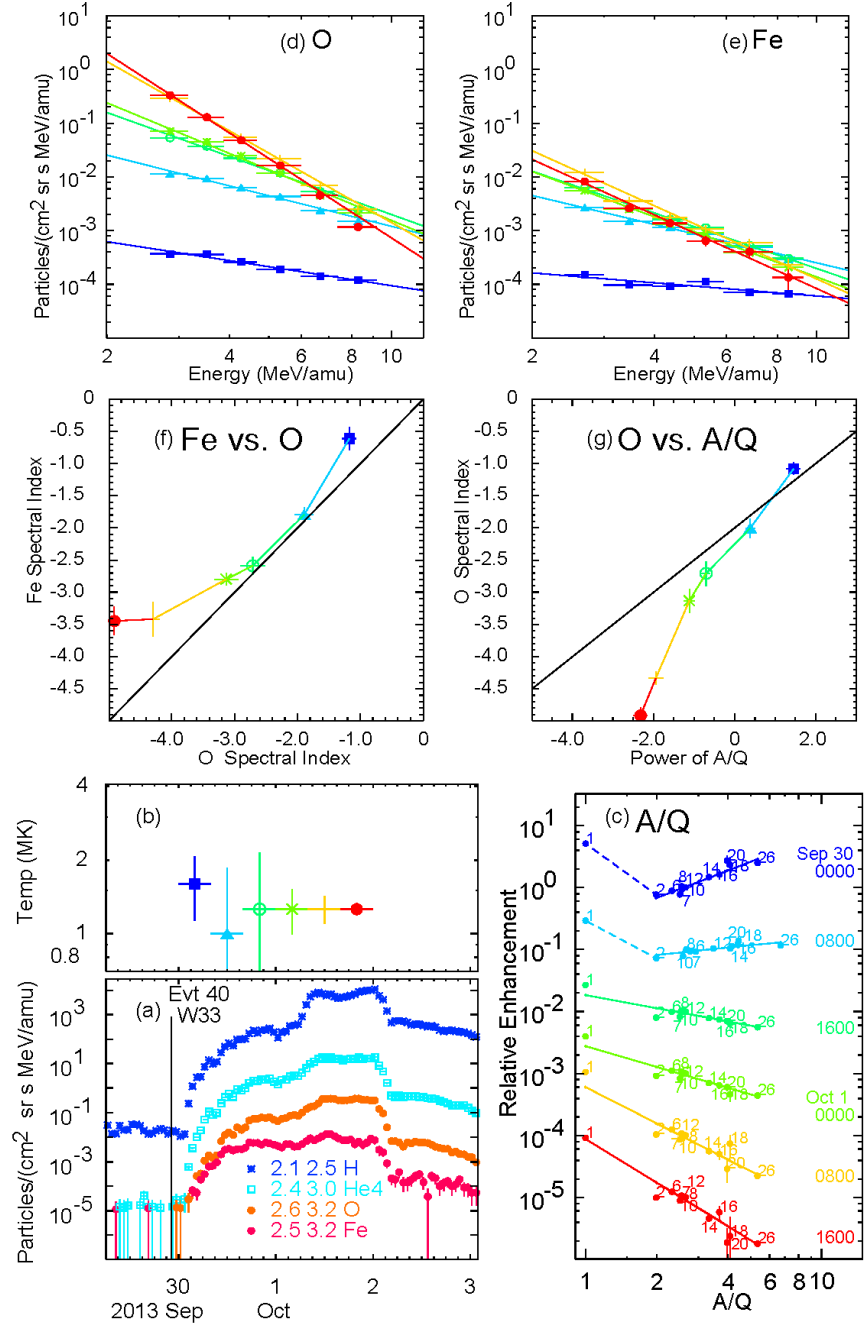

**Figure 4** (**a**) Intensities and (**b**) derived source temperatures are shown versus time for gradual SEP Event 40, of 29 September, 2013. Fits are shown for (**c**) enhancements of elements, listed by $Z$, versus $A/Q$, and for energy spectra of (**d**) O and (**e**) Fe. Correlation plots are shown for spectral indices of (**f**) Fe versus O and of (**g**) O versus $A/Q$. Colors for time intervals correspond in (**b**), (**c**), (**d**), (**e**), (**f**), and (**g**). In (**f**) the solid line is diagonal, $y = x$, in (**g**) it is $y = x/2 - 2$.

However, comparing smaller events that show almost no spectral variation with longitude or time, suggests that transport definitely dominates the behavior in these large events, once scattering by self-amplified waves becomes a factor. This wave generation and scattering is itself likely to be a strong function of longitude, varying with the streaming proton intensity. We will see that explanation of complex details of the time variations suggests that these events are indeed transport dominated.

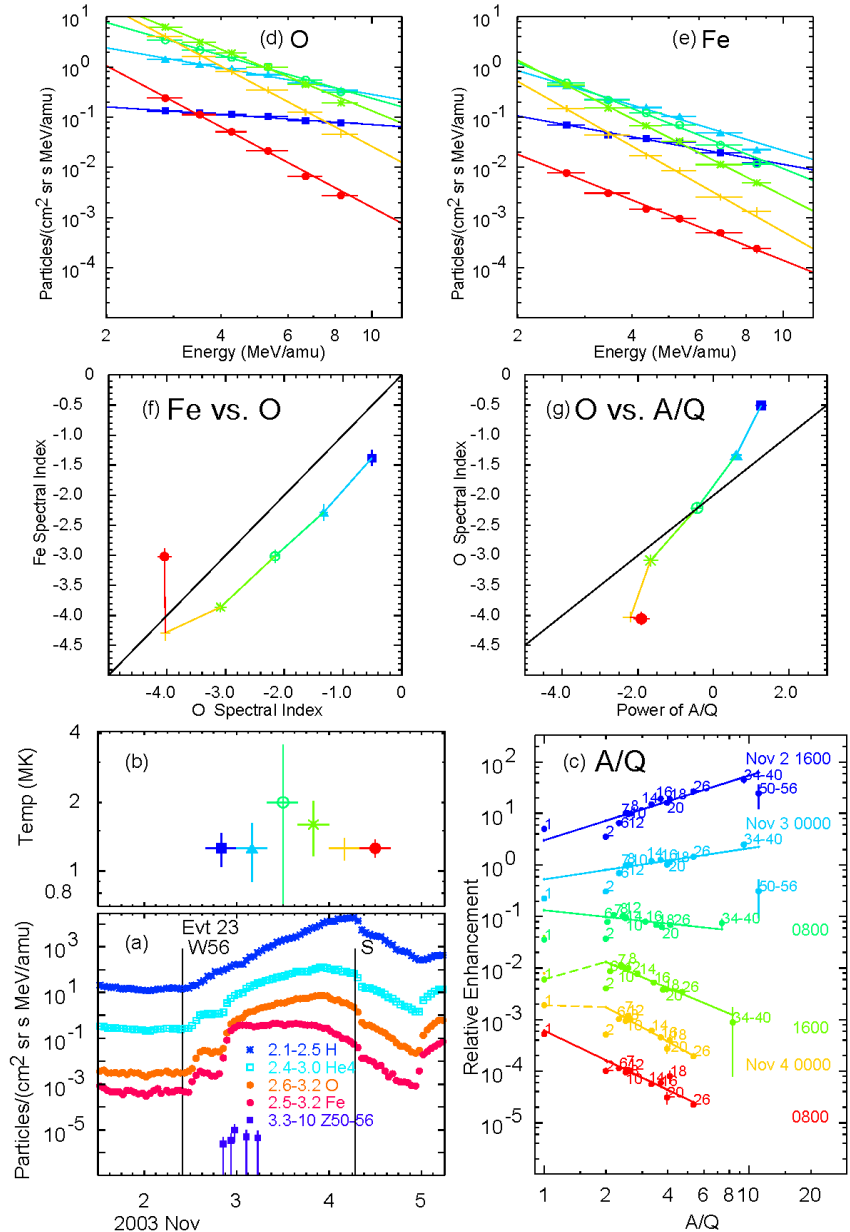

**Figure 5** (**a**) Intensities and (**b**) derived source temperatures are shown versus time for gradual SEP Event 23, of 2 November, 2003. Fits are shown for (**c**) enhancements of elements, listed by *Z*, versus *A/Q*, and for energy spectra of (**d**) O and (**e**) Fe. Correlation plots are shown for spectral indices of (**f**) Fe versus O and of (**g**) O versus *A/Q*. Colors for time intervals correspond in (**b**), (**c**), (**d**), (**e**), (**f**), and (**g**). In (**f**) the solid line is diagonal, y = x, in (**g**) it is y = x/2 – 2.

## 4. Large Gradual Events at the Streaming Limit

Intensities early in the largest gradual events can be limited by waves amplified by the high intensities of streaming SEP protons themselves, i.e. transport clearly dominates this behavior (Reames and Ng, 1998, 2010; Ng, Reames, and Tylka, 2012). This limit is also implicitly contained in the steady-state streaming models of Bell (1978a, 1978b) and Lee (1983, 2005). Some properties of these events include the time evolution of powers of *A/Q* that are positive upstream of the shock and negative downstream, shown by Reames (2020b).

Figure 6 shows the large SEP4 gradual event of 4 November 2001. This event also shows evolution of the power of *A/Q* from positive to negative values, but only in a narrow range, from +1 to -1. The spectral indices only vary from $E^{-1}$ to $E^{-2}$ across the shock and thus remain much harder later than in the event in Fig. 5. In fact, the region behind the shock appears to be a "reservoir" where magnetically trapped particle intensities decrease adiabatically with no change in spectral shape (Fig. 6f) as the volume of the magnetic "bottle" expands (e.g. Reames, 2013). Other large gradual events that show streaming limited behavior (Reames and Ng, 2010), e.g. 14 July, 2000 and 28 October, 2003, behave similarly, with flattened spectra and enhanced abundances early and a return near the line of correlated abundances and spectra behind the shock. Often they show the invariant spectra of a reservoir behind the extremely wide, fast CMEs that drive these events.

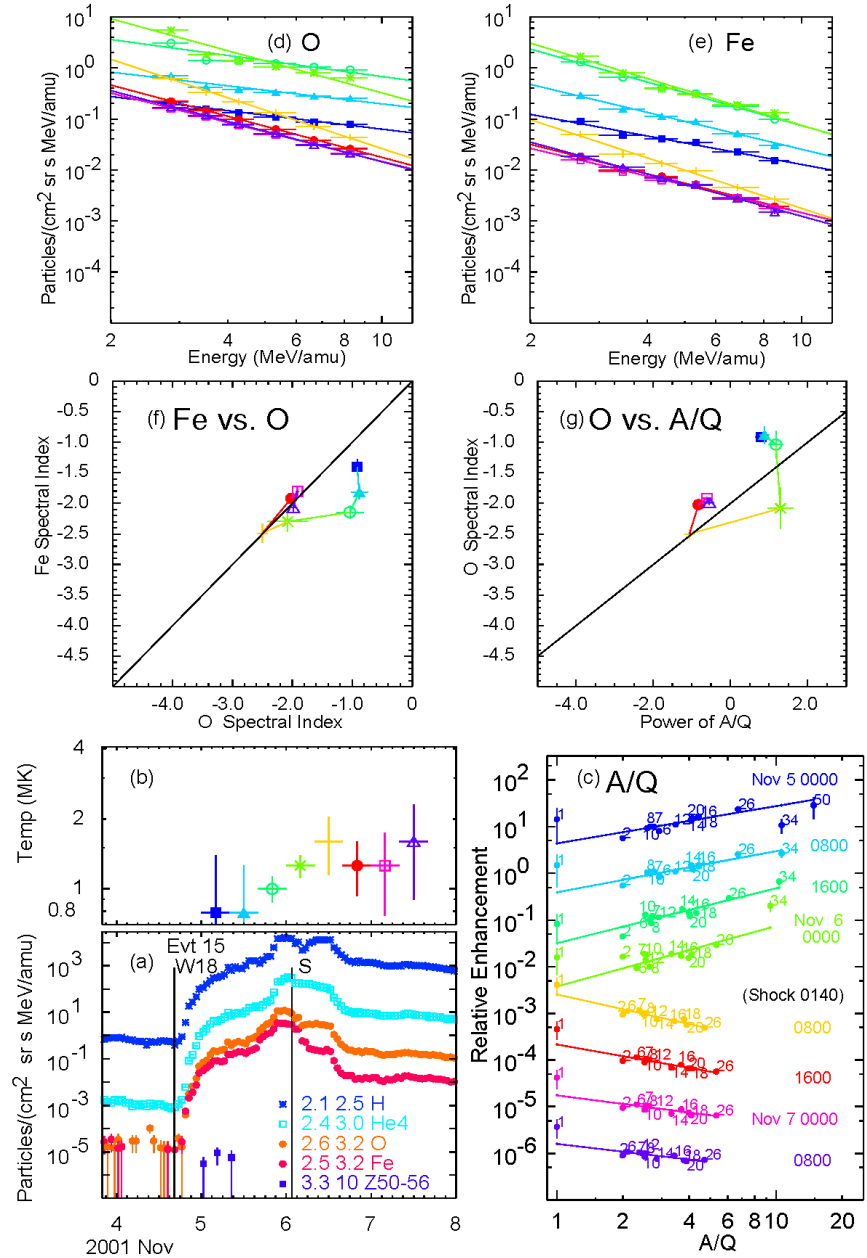

**Figure 6** (**a**) Intensities and (**b**) derived source temperatures are shown versus time for gradual SEP Event 15, of 4 November, 2001. Fits are shown for (**c**) enhancements of elements, listed by *Z*, versus *A/Q*, and for energy spectra of (**d**) O and (**e**) Fe. Correlation plots are shown for spectral indices of (**f**) Fe versus O and of (**g**) O versus *A/Q*. Colors for time intervals correspond in (**b**), (**c**), (**d**), (**e**), (**f**), and (**g**). In (**f**) the solid line is diagonal, y = x, in (**g**) it is y = x/2 − 2.

It is possible to study these large events with 2-hour time resolution to obtain better details on the tracks of the variations of the spectra and abundances. Figure 7 revisits the events of Figs. 5 and 6 with this improved resolution. The temperatures in Figs. 7b and 7f provide time sequences of color-coded symbols that can be used to follow the time evolution of the energy-spectral indices and powers of *A/Q* in the panels above. The number of intervals is too numerous to show each energy spectrum and fit of enhancement versus *A/Q*, but 8-hour versions of these spectra and fits are shown in Figs. 5 and 6.

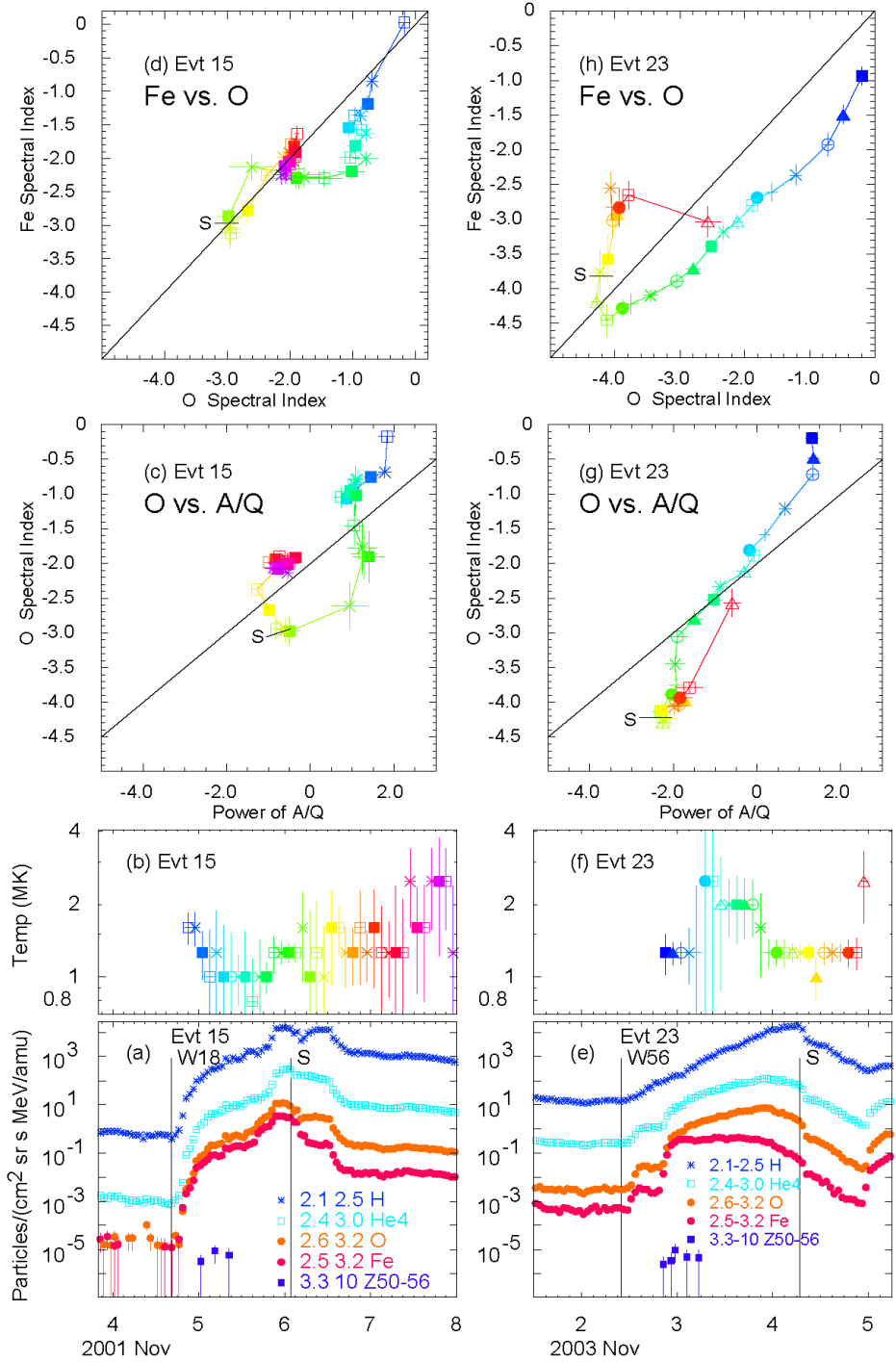

**Figure 7** revisits Event 15 of 4 November, 2001 from Fig. 6 on the *left*, and Event 23 of 2 November, 2003 from Fig. 5 on the *right* with 2-hour time resolution. Panels (**a**) and (**e**) show intensities of listed species and energies (in MeV amu$^{-1}$) versus time; (**b**) and (**f**) show derived temperatures versus time, providing time-tagged symbols and colors for the upper panels; (**c**) and (**g**) show time evolution of O spectral index versus power of *A/Q*; (**d**) and (**h**) show time evolution of Fe versus O spectral indices. S denotes the time of shock passage.

Both events in Fig. 7 are very large events.  Event 15 was among the examples with streaming-limited intensities and spectra early in the event (Reames and Ng, 2010), and both Events 15 and 23 are ground-level events (GLEs; Cliver, 2006; Gopalswamy et al., 2012).  Both are strongly transport-modified or even transport-limited events.

To compare spectra of Fe and O we must consider ion rigidities.  At $T \approx 1.3$ MK, our measurements of O span the range of about 180 – 360 MV and Fe spans about 370 – 700 MV.  The power of $A/Q$ spans from about 160 – 520 MV.   Thus, if rigidity spectra steepen at high rigidities while low-rigidity ions are retarded or trapped our Fe spectra will be steeper than our O spectra.  This is clearly seen in Fig. 7h, less so in Fig. 7d.

The steepest spectra occur near the time of shock passage, indicated by S.  This may seem surprising, since this is the particle source, but the weakened shock that arrives at 1 AU is accelerating only lower-energy ions.  Low-energy ions peak near the shock where they are detained, but higher energy ions pass through with no peak.  The steep shock spectra occur in our energy range.  The clustering of red points in Figs. 7c and 7d indicate the presence of a "reservoir" of magnetically trapped particles with adiabatically unchanging spectra (Reames, Kahler, and Ng, 1997; Reames, 2013, 2021).   These clusters of points return to near the expected lines in Fig. 7c and 7d.

In Fig. 7d, the spectra are extremely flat initially, reflecting the streaming-limited plateau (Reames and Ng, 2010; Reames, 2020b).  Then the Fe spectra steepen while O spectra do not since the break at energies above the flattening depends upon rigidity (see typical spectra in Fig 7f of Reames, 2020b or Fig. 3 in Reames and Ng, 2010).  Then the O spectrum begins to break and recover the same spectral index as Fe.

In Fig 7h, we note that the Fe spectrum is nearly one power steeper than the O spectrum all the way to the time of shock passage, and then it begins to reveal a hardening Fe spectrum as the low-rigidity ions, quasi-trapped near the shock, are carried past Earth.  The last time point, the red triangle, actually begins a new event that is not related.

The behavior of the evolution of the spectra and abundances in these large events is both interesting and complex, and not all of the features are well understood.  We can suggest narratives to explain the observed behavior, but quantitative theoretical models to test these ideas would be helpful.

## 5. Impulsively-Seeded Shock (SEP3) Events

So far, we have not yet considered those gradual events where the SEPs with $Z \geq 2$ are dominated by ions with $T > 2$ MK believed to be reaccelerated from pools of pre-enhanced impulsive suprathermal ions, the SEP3 events (Reames, 2020a). These events are identified by the strong heavy-ion enhancement throughout, the characteristic derived source temperature of ≈3 MK, and a large excess of protons relative to the extrapolated power-law extended from the enhancements of $Z > 2$ ions versus $A/Q$. The proton excess presumably comes because the shock can sample protons from the ambient coronal plasma but impulsive seed particles dominate the ions with $Z > 2$ (Reames, 2020a).

Figures 8 and 9 analyze examples of SEP3 events. Figure 8 shows spectra and abundances for the event of 4 August, 2011, and Fig. 9 shows data for the event of 14 November 1998. Both events show reasonably good agreement between the spectral indices of O and Fe in panel (f), in fact, the Fe spectra are slightly harder than the corresponding O spectra. Panel (g) for both figures shows that the power of $A/Q$ remains nearly constant at strongly positive values throughout the events. Thus the power of $A/Q$ is nearly constant with time and independent of the spectral indices for these events. This is not surprising since the power of $A/Q$ at $Z > 2$ is primarily determined by the ≈ 3 MK impulsive seed particles, born of magnetic reconnection in jets, but the spectral indices are determined by later acceleration at the shock. These are two completely different physical processes.

Apart from the unique abundances of SEP3 events, provided by their source, their shock-generated energy spectra vary little with time, making these events similar in behavior to small and moderate SEP4 events in this respect.

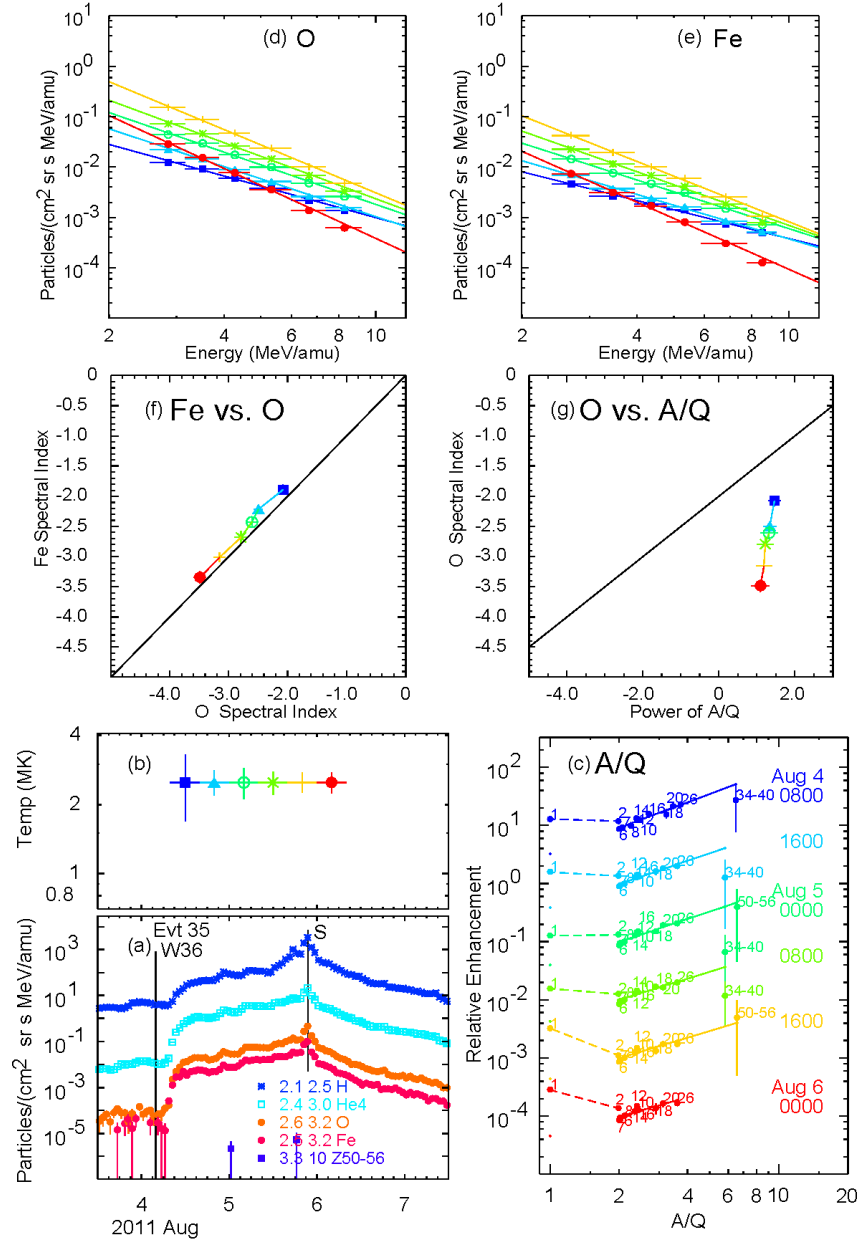

**Figure 8** (**a**) Intensities and (**b**) derived source temperatures are shown versus time for gradual SEP Event 35, of 4 August, 2011. Fits are shown for (**c**) enhancements of elements, listed by $Z$, versus $A/Q$, and for energy spectra of (**d**) O and (**e**) Fe. Correlation plots are shown for spectral indices of (**f**) Fe versus O and of (**g**) O versus $A/Q$. Colors for time intervals correspond in (**b**), (**c**), (**d**), (**e**), (**f**), and (**g**). In (**f**) the solid line is diagonal, $y = x$, in (**g**) it is $y = x/2 - 2$.

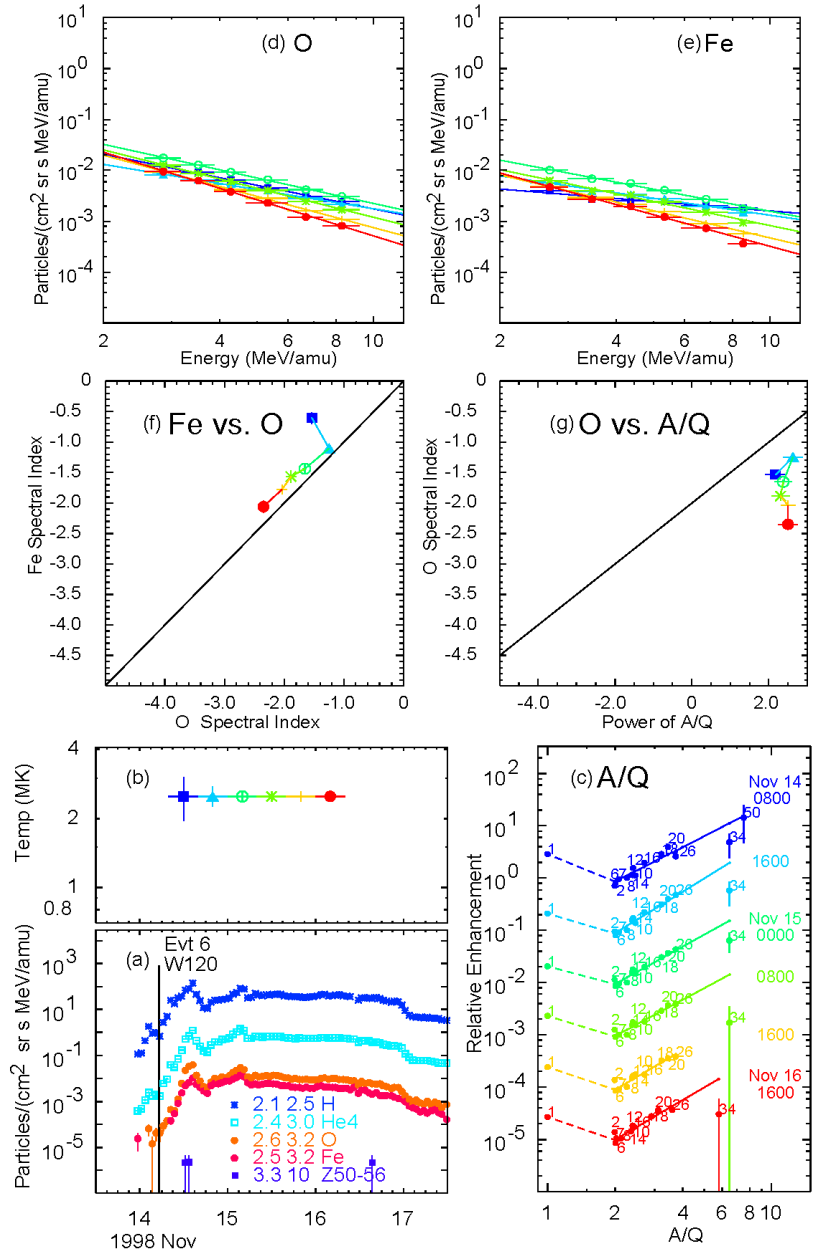

**Figure 9** (**a**) Intensities and (**b**) derived source temperatures are shown versus time for gradual SEP Event 15, of 14 November, 1998. Fits are shown for (**c**) enhancements of elements, listed by $Z$, versus $A/Q$, and for energy spectra of (**d**) O and (**e**) Fe. Correlation plots are shown for spectral indices of (**f**) Fe versus O and of (**g**) O versus $A/Q$. Colors for time intervals correspond in (**b**), (**c**), (**d**), (**e**), (**f**), and (**g**). In (**f**) the solid line is diagonal, y = x, in (**g**) it is y = x/2 – 2.

## 6. Spectral Indices and Powers of *A/Q*: Gradual Events

In Fig. 10 we examine the distribution of gradual SEP events in the space of spectral indices and spectral indices versus enhancement power laws $A/Q$. Rather than try to deduce some average behavior for each whole gradual event, we plot the properties of the first four 8-hour intervals in each event. The SEP3 events are well identified here by $T >$ 2 MK, they could just as easily have been identified by their proton excess.

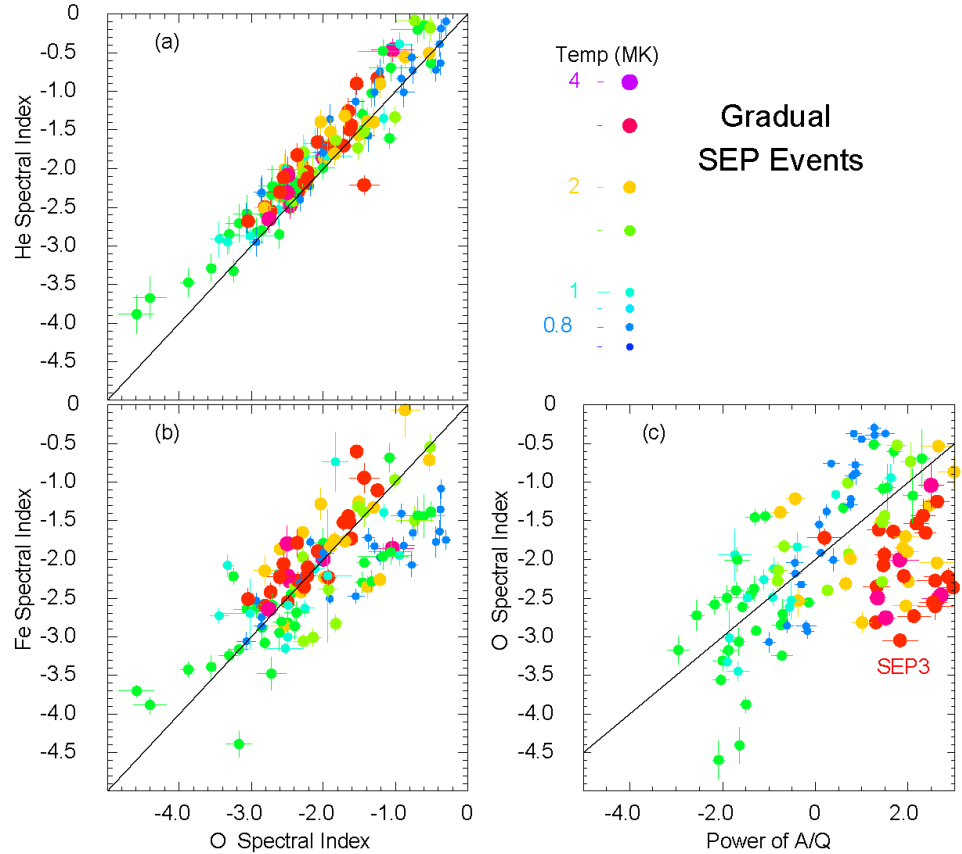

**Figure 10** Shown are (**a**) spectral indices of He versus O, (**b**) spectral indices of Fe versus O, and (**c**) spectral indices of O versus powers of *A/Q* for the first four 8-hour periods in the 45 gradual SEP events listed by Reames (2016). The size and color of each point is determined by the source plasma temperature *T* as shown by the scale. Events with *T* > 2 MK are SEP3-class events dominated by reaccelerated impulsive suprathermal ions with enhanced abundances already determined before shock acceleration. In (**a**) and (**b**) the solid line is diagonal, y = x, in (**c**) it is y = x/2 – 2.

Figure 10a shows that the spectra of He tend to be slightly harder than the corresponding O spectra for all types of events. In Fig. 10b, the Fe spectral indices show a somewhat larger variation, with Fe harder than O for the SEP3 events and Fe sometimes steeper than O for the remaining SEP4 events, on average. In Fig. 10c the SEP3 events are clustered at positive powers of *A/Q* and if these are removed, the SEP4 events are widely distributed along the line of expected correlation, but in some cases deflected from it by dislocations like those we saw in Fig. 7. The power of *A/Q* depends upon the reference abundances to define enhancements, but the spectral indices do not. The reference abundances for SEP3 events are actually impulsive suprathermal ion abundances, which may vary from event to event, but since we have assumed reference coronal abun-dances in all cases, the SEP3 events fall to the right of the line of expectation in Fig. 10c for a given spectral index of O.

## 7. Spectral Indices and Powers of *A/Q*: Impulsive Events

Figure 11 shows distributions of the 111 impulsive SEP events listed by Reames, Cliver and Kahler (2014a) in panels similar to those in Fig. 10.

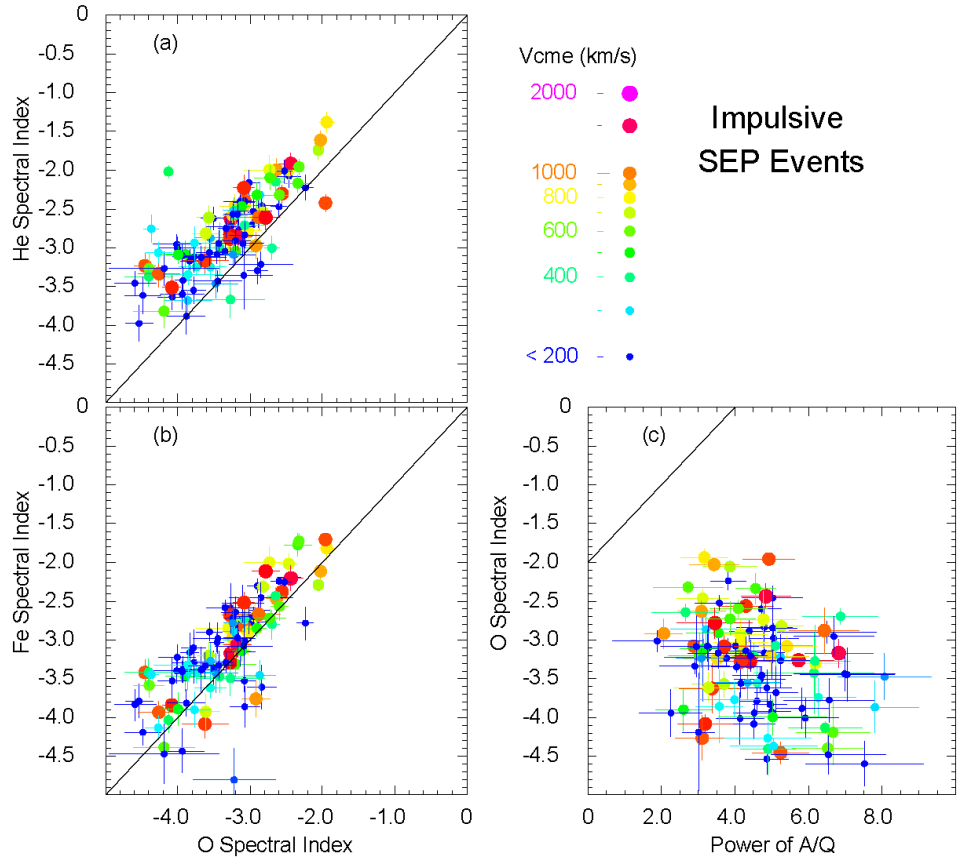

**Figure 11** Shown are (**a**) spectral indices of He versus O, (**b**) spectral indices of Fe versus O, and (**c**) spectral indices of O versus powers of *A/Q* for 111 SEP events (listed by Reames, Cliver, and Kahler, 2014a), the size and color of each point determined by the speed of an associated CME, if any. In (**a**) and (**b**) the solid line is diagonal, y = x, in (**c**) it is y = x/2 – 2.

For impulsive SEP events the source plasma temperatures are all ≈3 MK (Reames, Cliver, and Kahler, 2014b), so they convey no additional information; therefore, events have been colored by CME speeds instead. Note that values along the abscissa in Fig. 11c are much larger than those in Fig. 10c, so the points fall far from the values expected when spectra and abundances are determined together.  Spectra and abundances of all impulsive events are evidently determined independently.

Figure 11a shows that He spectra are harder than O spectra, but Fig. 11b shows that Fe spectra are also harder than O spectra, an interesting combination.  The spectral difference between Fe and O might be taken to suggest that the accumulation of this difference out to MeV energies has caused the enhancement in Fe/O, etc.  However, the correlation coefficient for the difference between Fe and O spectral indices versus the power of enhancement of $A/Q$ is only 0.26.  Furthermore, the average power of $A/Q$ for impulsive SEP events is 3.26 at $\approx$ 400 keV amu$^{-1}$ (Mason et al., 2004) and 3.64±0.15 at 3–5 MeV amu$^{-1}$ (Reames, Cliver, and Kahler, 2014).  These abundances do not have a significant variation with energy.

Using CME speed >500 km s$^{-1}$ as a proxy for the shock-reaccelerated SEP2 events, Fig. 10 suggests that their spectra are statistically somewhat harder than SEP1 events and their enhancements or powers of $A/Q$ are smaller.  As we have known, smaller impulsive SEP events tend to have more extreme enhancements.

## 8. Discussion and Conclusions

We should not be surprised that a process that hardens an energy spectrum, enhancing high-rigidity ions versus low-rigidity ions of the same species, might also systematically enhance high-rigidity Fe versus low-rigidity O, at a given velocity.  We have found that the SEP4 gradual events, in which shock waves accelerate ambient coronal plasma, often broadly tend to agree with the expected correlation between energy spectral indices and powers of abundance enhancement variation with $A/Q$, although with many interesting and complex variations.  Small and moderate gradual SEP events, where we seem to look directly at the shock source, show little variation in spectral indices or power-laws in $A/Q$ with time.  These events reflect the physics of the shock unmodified by transport.  Larger events show significant temporal variations that generally evolve away from the track of correlation, mostly because of scattering during transport.  Very large events have enhanced heavy-ion abundances and flattened energy spectra early and return to steeper spectra with correlated powers of $A/Q$ after shock passage; frequently the invariant spectra of a magnetic reservoir are seen behind the wide, fast CMEs that drive these events.

In essence the complex behavior of the large SEP4 events results from a breakdown of the power-law assumption. In the small and moderate SEP4 events a single power law spans the rigidity space from O at ≈180 – 360 MV up to Fe at ≈370 – 700 MV, so that O and Fe have similar power-law spectral indices. Transport has little effect. In the large SEP4 events, wave generation during transport affects low-rigidity O much more than high-rigidity Fe, creating different spectral indices that also vary strongly with time. We can fit power laws locally but not over the full rigidity and time span – a breakdown. As the events evolve outward, the time variations mix high- and low-rigidity ions modified by different wave spectra in different places at different times.

Would we still find an extensive power law close to the shock in large events, as we see for smaller events? Perhaps the more efficient trapping of low-rigidity ions near the shock would have feedback that leads to more rapid acceleration to higher energies. Perhaps this additional trapping, produced by intense streaming protons, is also evidence of the presence of high-energy protons that can produce a GLE. The highest energy Fe resonates with waves produced by streaming protons above 230 MeV, yet the spectra and abundances we consider are all in power-law regions of the shock spectra, below 10 MeV amu$^{-1}$ and below any disruptive spectral breaks.

As we might expect, the SEP2 and SEP3 events, where shocks are seeded by pre-accelerated impulsive ions, show no systematic correlation between spectra and abundances. Shock acceleration, which determines the spectra, probably modifies or decrease the power-law abundances somewhat, but those strong positive abundance enhancements primarily survive from the seed population. We cannot easily renormalize the abundances to reference seed populations with contributions from both impulsive and ambient-coronal components which vary from event to event.

It is also true that there is no apparent correlation between spectra and abundances for the SEP1 events. For these events, any accompanying CME is not fast enough to drive a shock wave that can reaccelerate ions to MeV energies. If these events involve a single acceleration process (e.g. Drake et al., 2009), the abundances must be determined separately from the spectra. Probably the abundance enhancements are determined prior to the actual acceleration.

Correlations between Spectra and Abundances of SEPs                D. V. ReamesIon scattering during transport is a strongly rigidity-dependent process. If low-rigidity ions are preferentially scattered and trapped, that will flatten early spectra and also retard O relative to Fe, increasing Fe/O early, a positive power of *A/Q*. Transport imposes a new rigidity dependence that does not maintain the same relationship between spectra and abundances. However, positive powers of *A/Q* for the cooler coronal-sourced ($T < 2$ MK) ions early in events surely come from preferential scattering and retardation of low-rigidity ions during transport, trapping them near the shock source. Complex transport-dominated features control and often distort the behavior of large gradual SEP4 events.

We have found that SEP4 events with $E^y$ spectra and $(A/Q)^x$ abundances have an underlying relationship of $y = x/2 - 2$, i.e. a power of velocity of $x - 4$. These relations were determined from fitting the data. Presumably, this newly-discovered relationship results from the basic physics of ion selection by shocks, but its connection to first principles is not yet known. What is the origin of the -2 or -4? This is a new way to attack the "injection problem" (e.g. Zank et al. 2001).

Presumably, modeling could associate the spectra and abundances with shock compression ratios, etc. Are there also measureable features of the correlated spectra and abundances that arise from properties and structure of the underlying shock waves? Current SEP-acceleration models treat injection of ions as adjustable parameters, if at all, yet we know the ion abundances in the source plasma very well. We need theories and models that address both spectra and abundances.

## Disclosure of Potential Conflicts of Interest

The author declares he has no conflicts of interest.

## References

Afanasiev, A., Battarbee, M., Vainio, R.: 2016, Self-consistent Monte Carlo simulations of proton acceleration in coronal shocks: Effect of anisotropic pitch-angle scattering of particles, *Astron. Astrophys.* **584,** 81 doi: 10.1051/0004-6361/201526750

Bell, A.R.: 1978a, The acceleration of cosmic rays in shock fronts. I, *Mon. Not. Roy. Astron. Soc*. **182**, 147 doi: 10.1093/mnras/182.2.14724